\documentclass[reprint,superscriptaddress,amsmath,amssymb,aps,floatfix]{revtex4-2}
\usepackage{graphicx}
\usepackage{color}

\usepackage{dcolumn}
\graphicspath{{./Figures/}}
\usepackage{upgreek}
\usepackage{gensymb}
\usepackage{float}
\usepackage{bm}
\usepackage{xcolor}
\usepackage{physics}
\usepackage{longtable}
\usepackage{comment}

\usepackage{xr}
\makeatletter
\newcommand*{\addFileDependency}[1]{
	\typeout{(#1)}
	\@addtofilelist{#1}
	\IfFileExists{#1}{}{\typeout{No file #1.}}
}
\makeatother

\newcommand*{\myexternaldocument}[1]{
	\externaldocument{#1}
	\addFileDependency{#1.tex}
	\addFileDependency{#1.aux}
}

\myexternaldocument{ME_CFB_ver1_supps}

\begin{document}
	
\title{Acoustoelectric non-local spin wave power detector for studying magnon-phonon coupling}
\author{Hiroki Matsumoto}
\affiliation{Department of Physics, The University of Tokyo, Tokyo 113-0033, Japan}
\affiliation{Institute for Chemical Research, Kyoto University, Kyoto 611-0011, Japan}

\author{Yasuhiro Todaka}
\affiliation{Department of Physics, The University of Tokyo, Tokyo 113-0033, Japan}

\author{Takuya Kawada}
\affiliation{Department of Physics, The University of Tokyo, Tokyo 113-0033, Japan}
\affiliation{Department of Physics, Osaka University, Osaka 560-0043, Japan}

\author{Masashi Kawaguchi}
\affiliation{Department of Physics, The University of Tokyo, Tokyo 113-0033, Japan}

\author{Masamitsu Hayashi}
\affiliation{Department of Physics, The University of Tokyo, Tokyo 113-0033, Japan}
\affiliation{Trans-scale quantum science institute, The University of Tokyo, Tokyo 113-0033, Japan}

\newif\iffigure
\figurefalse
\figuretrue

\date{\today}

\begin{abstract}
We have developed a simple detection scheme to study spin waves excited by surface acoustic wave (SAW) in ferromagnetic thin films. 
Metallic antennas made of Ta and a ferromagnetic element are placed along the SAW propagation path.
The SAW excites spin waves in the ferromagnetic element and induces acoustoelectric current in the antennas, which are detected as a DC voltage.
The DC voltage takes an extremum at the spin wave resonance condition, which demonstrates that the antenna acts as a non-local spin wave detector. 
The antennas placed before and after the ferromagnetic element along the SAW propagation path can probe spin wave excitation from reflected and transmitted SAWs, respectively.
Interestingly, we find characteristics of spin wave excitations that are different for the reflected and transmitted SAWs: the former excites spin waves with larger frequency with broader linewidth compared to the latter.
The results show that the antennas act as a non-local spin wave power detector and can be used to map out the spin wave spectra in a unique way, providing insights into the magnon-phonon coupling in magnetic nanostructures fabricated on phononic SAW devices.
\end{abstract}

\maketitle

Surface acoustic wave (SAW)\cite{Rayleigh1885,delsing2019jpd} is an acoustic phonon that can be readily excited using piezoelectric substrates.
A widely used approach to generating SAW is to pattern a comb-shaped electrode, often referred to as an interdigital transducer (IDT), on a piezoelectric substrate and apply a microwave electrical signal to the electrode\cite{White1965}.
The electric field applied to the IDT is converted to an oscillating strain due to the substrate piezoelectricity.
The strain is launched as an acoustic wave along the substrate surface, propagating a distance of hundreds of micrometers to millimeters\cite{Casals2020}.
One may deposit a thin film on the path to which the SAW travels, thus allowing excitation of acoustic phonons in the film.

Owing to its high coherence, SAW has been used to study the interaction between acoustic phonons and other degrees of freedom in solids\cite{shao2019prap,manenti2016prb}.
Among them, coupling with spin waves\cite{kruglyak2010jpd} in magnetic thin films are gaining attraction since their eigenenergy and wavelength can be close to those of SAW\cite{Weiler2011,Dreher2012,Thevenard2014,Kraimia2020,Sasaki2021}.
Recent studies have examined the strength of the coupling between SAWs and spin waves\cite{Hatanaka2022, asano2023prb, hwang2024prl, matsumoto2023arXiv}.
The coupling often manifests itself as extra absorption of SAW power that travels across a magnetic element. 
Studies have also shown that transmission of SAW becomes non-reciprocal when it couples to spin waves\cite{Sasaki2017, Tateno2020, HernandezMinguez2020, Shah2020, Xu2020, Kuess2020, Matsumoto2022}.

The spin wave-SAW coupling, or magnon-phonon coupling in general, may cause non-linear effects in the SAW and spin waves characteristics\cite{alekhin2023nanolett}. 
In analogy to non-linear optics, higher harmonic generation, parametric amplification and parametric down conversion are known to arise due to non-linearity.
It is, however, difficult to probe such non-linear effects unless one relies on sophisticated time- and space-resolved magnetic or atomic imaging techniques (e.g. magneto-optical Kerr effect\cite{kuszewski2018prap,Kraimia2020}, Brillouin light scattering\cite{Zhao2021, babu2021nanolett,Geilen2022}, photoemission electron microscopy\cite{Casals2020} and pulse induced microwave impedance microscopy\cite{nii2023prap}).

\begin{figure*}[bt]
	\begin{tabular}{cc}
		\begin{minipage}{0.47\hsize}
			\includegraphics[scale=0.45]{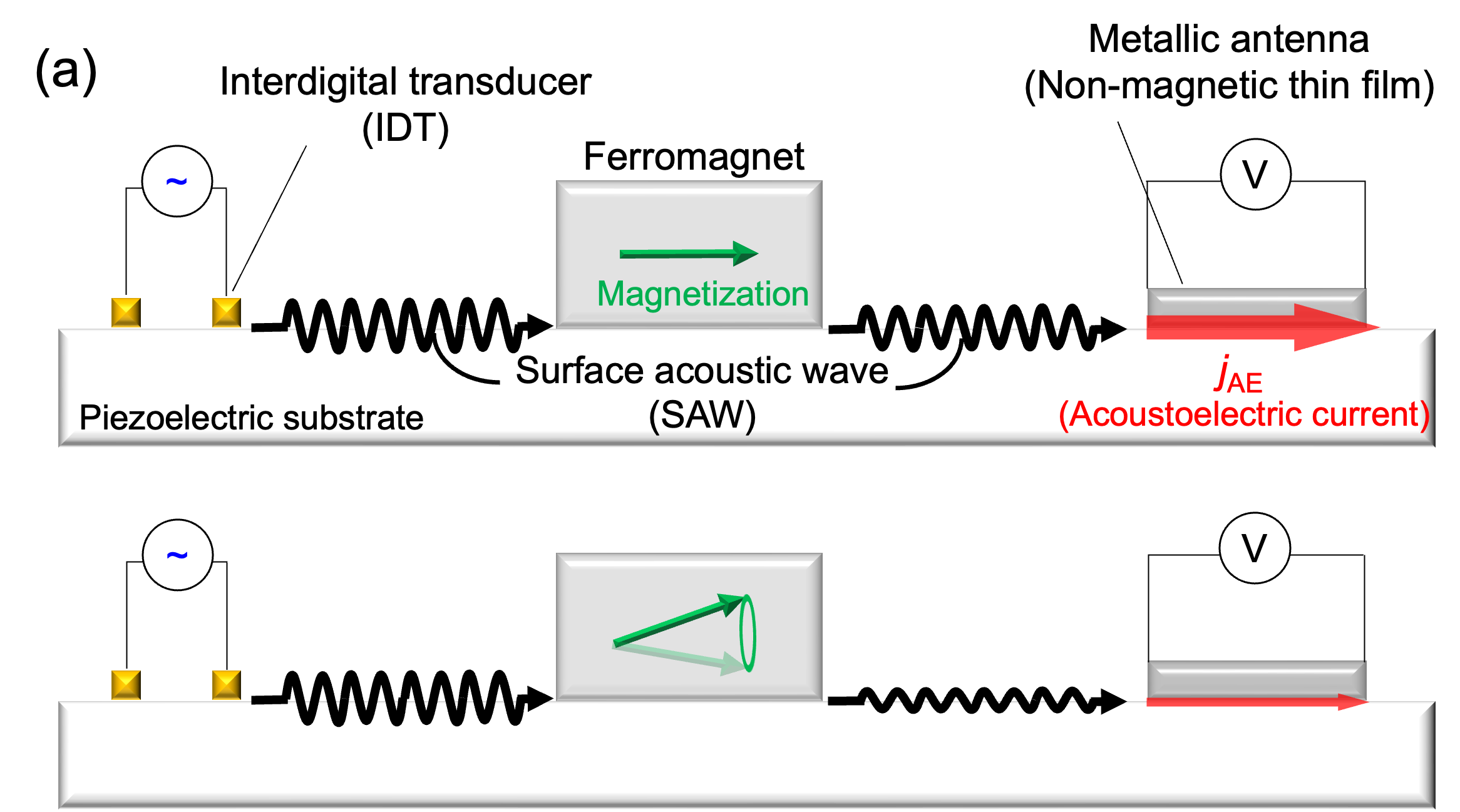}
		\end{minipage}
		\begin{minipage}{0.51\hsize}
			\includegraphics[scale=0.45]{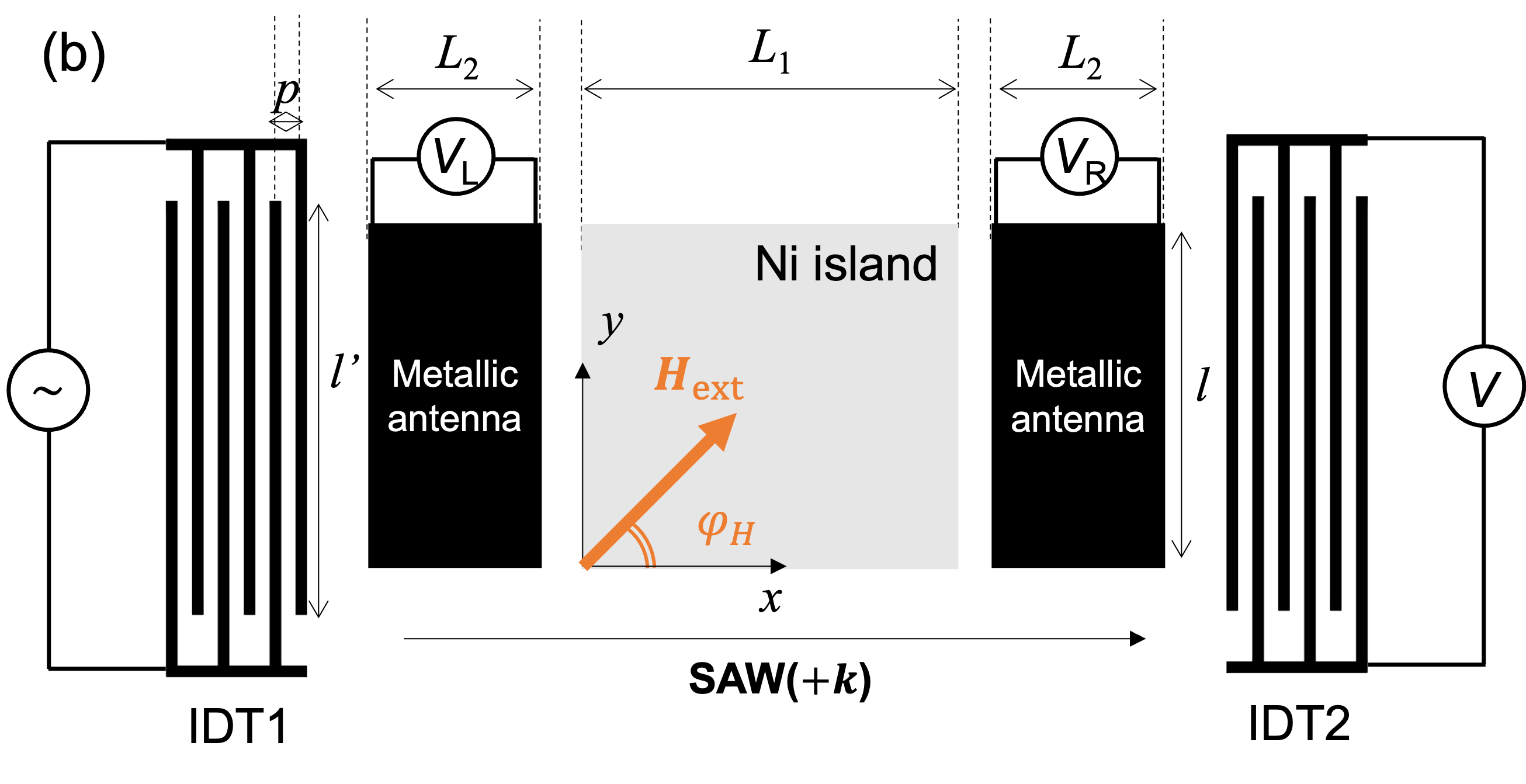}
		\end{minipage}
	\end{tabular}
	\caption{(a) Schematic illustration of the detection of SAW-driven spin waves by voltage measurement. SAW is excited at the IDT and propagates to the ferromagnetic thin film. The SAW generates acoustoelectric current in the metallic antenna, which is detected as DC voltage. When the SAW frequency matches the spin wave resonance condition, SAW power is absorbed by excitation of spin waves, which results in the reduction of the acoustoelectric current.  (b) Schematic illustration of the experimental setup. The island including 10 nm-thick Ni layer is located at the center of the delay line and a pair of metallic antenna (composed of Ta/MgO/Ta) are placed on both sides of the island. We apply rf power to IDT1 (IDT2) to excite SAW along $+k$ ($-k$). The device geometry are $p$ = 8 $\mu$m, $l$ = 400 $\mu$m, $l'$ = 450 $\mu$m,  $L_{1}$ = 450 $\mu$m and $L_{2}$ = 200 $\mu$m.
		\label{fig:setup}
	}
\end{figure*}

Here we show a simple approach to detecting SAW-induced spin waves via voltage measurements.
The concept of the detection scheme is illustrated in Fig.~\ref{fig:setup}(a). 
A metallic antenna, made of a Ta thin film, and a ferromagnetic island are patterned on a piezoelectric substrate along the path where the SAW travels.
The SAW excites spin waves in the ferromagnetic island under the resonance condition.
Simultaneously, the SAW induces acoustoelectric current $j_\mathrm{AE}$\cite{Parmenter1953,Weinreich1957} in the antennas, which can be detected as a DC voltage. 
Previous studies have shown that $j_\mathrm{AE}$ is proportional to the SAW power $P_\mathrm{SAW}$\cite{Miseikis2012, Poole2015, Kawada2019}.
The DC voltage takes a minimum at the spin wave resonance condition, which can be accounted for by SAW power absorption to the ferromagnetic island via spin wave excitation.
We show that the antennas placed before and after the ferromagnetic island along the SAW propagation path can probe, respectively, spin waves excited from SAWs reflected and transmitted at the island.

\par
Figure~\ref{fig:setup}(b) shows a schematic illustration of the experimental setup.
We fabricate a pair of IDTs (IDT1 and IDT2) composed of Ta(5)/Cu(50)/Pt(3) (units in nanometers) on a Y+128$^\mathrm{o}$-cut lithium niobate (LiNbO$_{3}$) substrate by radio frequency (rf) magnetron sputtering and a liftoff process.
The width and the gap of the IDT fingers are set to 2 $\mu$m.
The delay line (i.e. the path along IDT1 to IDT2) is patterned along the crystalline $X$-axis of LiNbO$_{3}$, which coincides with the $x$-axis shown in Fig.~\ref{fig:setup}(b).
$+k$ ($-k$) is defined as the propagation direction of SAW from IDT1 to IDT2 (IDT2 to IDT1) along the $x$-axis.
Upon forming the IDTs, a ferromagnetic Ni island made of Ni(10)/Cu(3)/MgO(2)/Ta(1) is placed at the center of the delay line.
The MgO(2)/Ta(1) serves as a capping layer.
To avoid oxidation of Ni via sputtering of MgO, the Cu(3) layer is inserted. 
The size of the island is $450 \times 400$ $\mu$m$^2$.
Subsequently, a pair of metal antennas composed of Ta(3)/MgO(2)/Ta(1) is formed on both sides of the Ni island.
Electrodes made of Ta(5)/Cu(50)/Pt(3) are attached to the antennas.
All films (IDTs, Ni island, antennas and the electrodes) are made by rf magnetron sputtering.
Details of the device geometry are described in Fig.~\ref{fig:setup}(b).
An external in-plane magnetic field with magnitude $\mu_{0}H_\mathrm{ext}$ is applied during the measurements.
The field angle $\varphi_\mathrm{H}$ is defined as the relative angle between the magnetic field and $+x$.
All measurements are performed at room temperature.

First, we characterize the device without exciting any spin waves in the ferromagnetic island.
A vector network analyzer (VNA) is used to evaluate the $S$-parameters.
Port $i$ ($i = 1,2$) of the VNA is connected to IDT$i$. 
An rf electrical signal is supplied to port $i$ and the transmitted electrical signal is measured at port $j$ to obtain the scattering matrix $S_{ji}$.
We find the fundamental mode (at $\sim$0.48 GHz) and the 5th harmonic mode ($\sim$2.35 GHz) associated with the Reyleigh SAW of the LiNbO$_{3}$ substrate appear in the $S_{12}$ and $S_{21}$ spectra measurements.
In the experiments described below, we use the 5th harmonic mode since its frequency is sufficiently large to excite spin waves in Ni\cite{Weiler2011, Dreher2012}.
$f_\mathrm{SAW}$ ($\sim$2.35 GHz) is defined as the SAW resonant frequency hereafter. 
During the spectrum measurements, a magnetic field of $\mu_{0}H_\mathrm{ext} = 110$ mT ($\varphi_\mathrm{H}$ = 90$^\mathrm{o}$) is applied to the device. 
The magnitude of the field is chosen such that spin wave excitation in the Ni island is suppressed in the frequency range studied.

\begin{figure}[bt]
		\centering
		\includegraphics[scale=0.5]{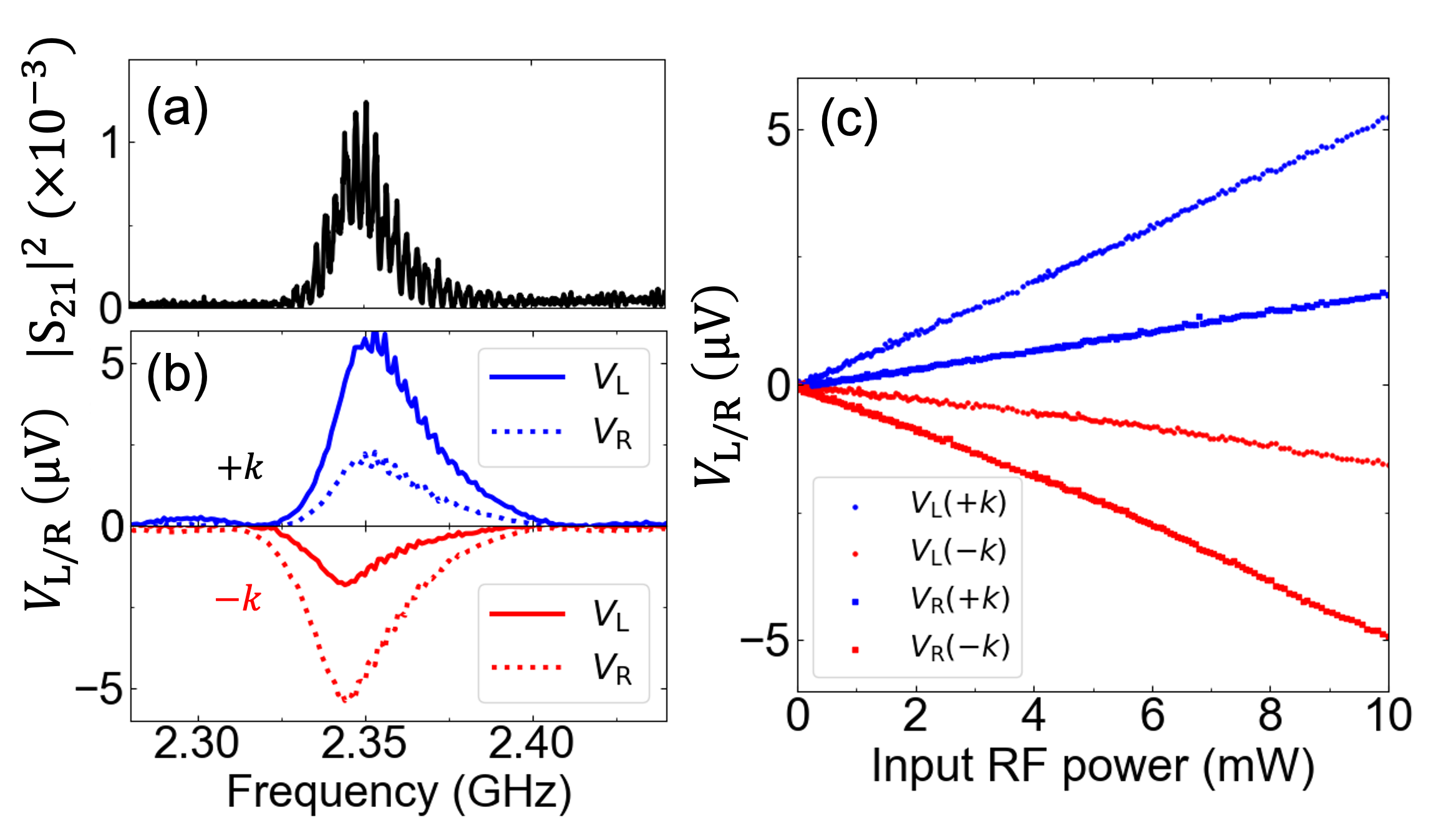}
	\caption{(a, b) Input rf frequency dependence of $|S_{21}|^{2}$ (a) and the voltages $V_\mathrm{L}$ and $V_\mathrm{R}$ of the antennas placed after the Ni island (b). 
	The rf power applied to the IDT is 10 mW. (c) Input rf power dependence of $V_\mathrm{L}$ and $V_\mathrm{R}$ at $f = f_\mathrm{SAW} \sim 2.35$ GHz. A magnetic field of $\mu_{0}H_\mathrm{ext}$ = 110 mT is applied along the $y$-axis ($\varphi_\mathrm{H} = 90^\mathrm{o}$) during the measurements.
	\label{fig:aec}
	}
\end{figure}

Figure~\ref{fig:aec}(a) shows the input frequency dependence of $|S_{21}|^{2}$ around $f_\mathrm{SAW}$.
Here we show the square of the transmission amplitude, which is proportional to the transmitted power.
$|S_{21}|^{2}$ takes a maximum at $f_\mathrm{SAW}$.
Multiple sub-peaks found in the spectra near $f_\mathrm{SAW}$ result from multiple reflection echo between the IDTs\cite{Kondoh2008}.
The response of the DC voltage across the metallic antennas is shown in Fig.~\ref{fig:aec}(b).
An rf signal source is connected to one of the IDTs: SAW travels along $+k$ ($-k$) when the source is connected to IDT1 (IDT2).
From the source, an amplitude modulated rf signal is applied to the IDT, which excites an amplitude modulated SAW.
The voltage across the antenna is measured using a lock-in amplifier.
The frequency and phase of the lock-in amplifier are locked to those of the rf signal source.
The voltage measured at the left and right antennas are defined as $V_\mathrm{L}$ and $V_\mathrm{R}$, respectively: see Fig.~\ref{fig:setup}(b) for the details of the configuration.
In Fig.~\ref{fig:aec}(b), the solid (dotted) lines show $V_\mathrm{L}$ ($V_\mathrm{R}$) when the power of the rf signal source is set to 10 mW.
The top and bottom panels show measurement results when the SAW travels along $+k$ and $-k$, respectively.

As evident, the magnitude of the voltage is enhanced when the rf signal source frequency is matched to $f_\mathrm{SAW}$. 
When the SAW travels along $+k$, we find a positive voltage both for $V_\mathrm{L}$ and $V_\mathrm{R}$ whereas the sign of the voltages changes when the SAW propagation direction is reversed.
This is consistent with the notion that the observed voltage is induced by the acoustoelectric current\cite{Miseikis2012,Kawada2019,Kawada2021}.
Note that the magnitude of the peaks differs depending on the SAW propagation direction and the probe position.
For example, the peak height of $V_\mathrm{L}$ is larger than that of $V_\mathrm{R}$ when SAW propagates along $+k$ and vice versa for $-k$.
The results show that the peak voltage of the antenna placed in the SAW propagation path before the Ni island is larger than that placed after the island.
We consider this is simply caused by mechanical damping/electromagnetic absorption of the SAW power at the Ni island.
Note that there is a slight difference in the extremum position (in frequency) of the voltage peaks when the SAW moves along $+k$ and  $-k$. 
We infer this is caused by unintended difference in the geometry (e.g. pitch size) of IDT1 and IDT2.
The rf power dependence of $V_\mathrm{L}$ and $V_\mathrm{R}$ are shown in Fig.~\ref{fig:aec}(c).
Clearly, $V_\mathrm{L}$ and $V_\mathrm{R}$ both linearly scale with the power, which corroborates the assumption that the detected voltages are due to the acoustoelectric current\cite{Miseikis2012, Poole2015, Kawada2019}.

\begin{figure}[t]
	\centering
	\includegraphics[scale=0.5]{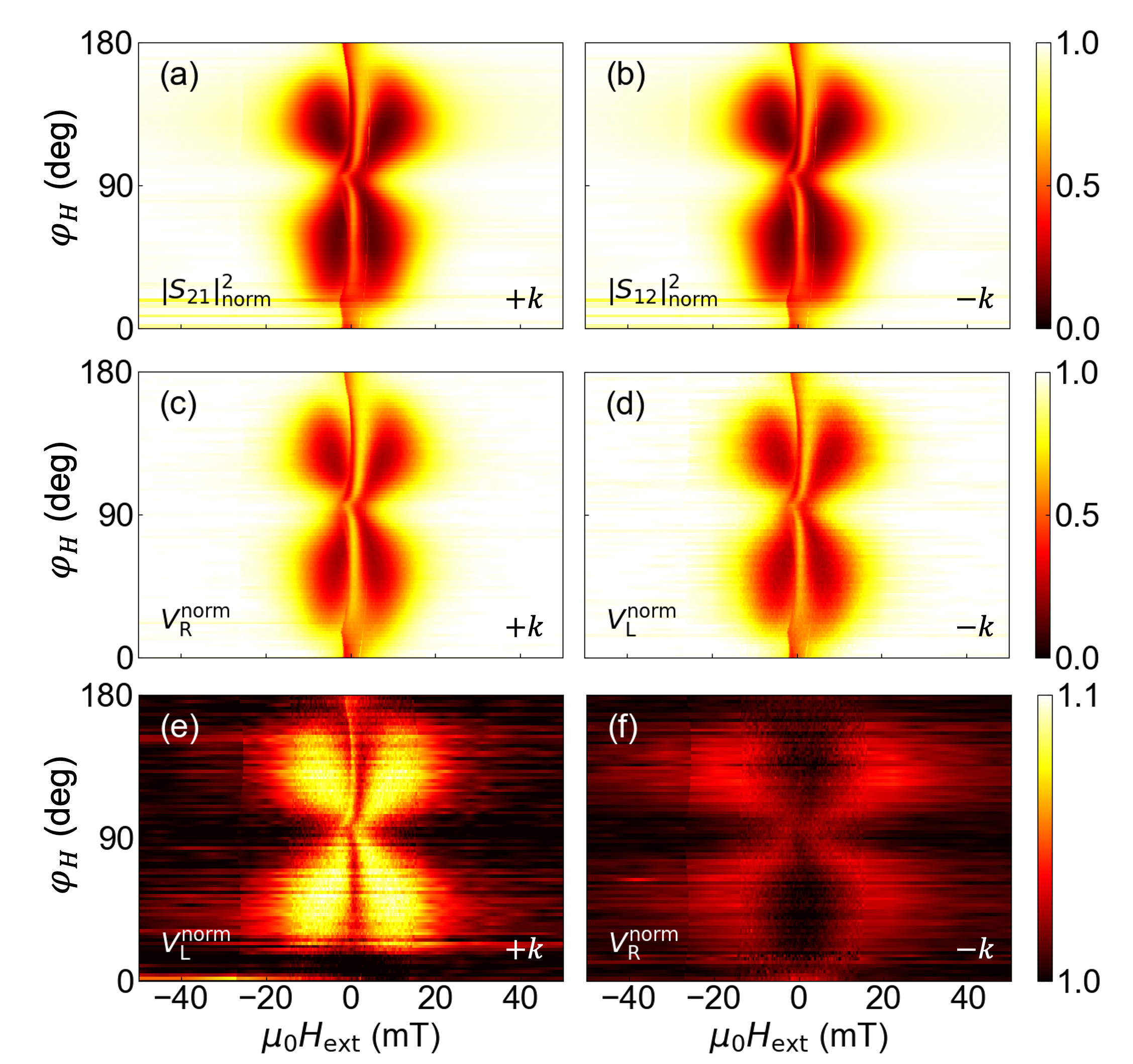}
	\caption{(a-d) Normalized transmitted power $|S_{21}|_\mathrm{norm}^{2}$ (a) and $|S_{12}|_\mathrm{norm}^{2}$ (b) and the corresponding normalized antenna voltages $V_\mathrm{R}^\mathrm{norm}$ (c) and $V_\mathrm{L}^\mathrm{norm}$ (d), obtained from the antennas placed after the Ni island, plotted as a function of the external magnetic field $\mu_{0}H_\mathrm{ext}$ and its orientation $\varphi_\mathrm{H}$. (e,f) Normalized antenna voltages $V_\mathrm{L}^\mathrm{norm}$ and $V_\mathrm{R}^\mathrm{norm}$, obtained from the antennas placed before after the Ni island, vs. $\mu_{0}H_\mathrm{ext}$ and $\varphi_\mathrm{H}$. An rf signal is applied to IDT1 (IDT2) and the SAW propagates along $+k$ ($-k$) for (a,c,e) ((b,d,f)). Input rf power is fixed to 10 mW.
		\label{fig:color}
	}
\end{figure}

Next, we vary the magnitude and direction of the in-plane magnetic field so as to excite spin waves in the Ni island.
Here we compare the $S$-parameter measurements using the VNA and the voltages at the antennas. 
In Figs.~\ref{fig:color}(a) and \ref{fig:color}(b), we show the normalized SAW transmission power, $|S_{21}|^{2}_\mathrm{norm}$ and $|S_{12}|^{2}_\mathrm{norm}$, plotted as a function of $\mu_{0}H_\mathrm{ext}$ and $\varphi_\mathrm{H}$.
The normalized power $|S_{ij}|^{2}_\mathrm{norm}$ ($i,j=1,2$) represents $|S_{ij}|^{2}$ obtained under a given field ($\mu_{0}H_\mathrm{ext}$, $\varphi_\mathrm{H}$) normalized by $|S_{ij}|^{2}$ measured at ($110$ mT, $\varphi_\mathrm{H}$).
For a fixed field angle, both $|S_{21}|^{2}_\mathrm{norm}$ and $|S_{12}|^{2}_\mathrm{norm}$ take a minimum at small $|\mu_{0}H_\mathrm{ext}|$.
Using the Kittel formula\cite{Kittel1948} with negligible magnetic anisotropy, we estimate the ferromagnetic resonance field $\mu_0 H_\mathrm{res}$ at $f_\mathrm{SAW}$ to be $\sim$ 10 mT.
The field at which the transmitted power takes a minimum is roughly equal to $\mu_0 H_\mathrm{res}$ at the frequency studied, indicating that the SAW power is absorbed by spin waves excited in the Ni island. 

The corresponding normalized DC voltages ($V_\mathrm{L}^\mathrm{norm}$ and $V_\mathrm{R}^\mathrm{norm}$) obtained at the antennas are shown in Figs.~\ref{fig:color}(c) and \ref{fig:color}(d), respectively.
Similar to $|S_{ij}|^{2}_\mathrm{norm}$, $V_\mathrm{L(R)}^\mathrm{norm}$ is defined as the voltage $V_\mathrm{L(R)}$ measured at a given field ($\mu_{0}H_\mathrm{ext}$, $\varphi_\mathrm{H}$) normalized by $V_\mathrm{L(R)}$ measured at ($110$ mT, $\varphi_\mathrm{H}$).
Here we show the normalized voltage of the antennas placed on a SAW propagation path after the Ni island ($V_\mathrm{R}^\mathrm{norm}$ for $+k$ and $V_\mathrm{L}^\mathrm{norm}$ for $-k$).
Thus the SAWs that travel and reach the antennas carry information of the spin waves excited at the Ni island.
As evident, the results resemble that of scattering matrix measurements (Fig.~\ref{fig:color}(a,b)).
These results show that the antenna serves as a non-local probe of SAW induced spin wave excitation in ferromagnetic elements.

In Figs.~\ref{fig:color}(e) and \ref{fig:color}(f), we show the normalized voltage of the antennas placed on the SAW propagation path before the Ni island ($V_\mathrm{L}^\mathrm{norm}$ for $+k$ and $V_\mathrm{R}^\mathrm{norm}$ for $-k$).
Here we expect no apparent signal from the spin waves at the Ni island if one considers SAWs that had traveled from the IDT to the antenna.
However, we find clear contrast in the contour plots in Fig.~\ref{fig:color}(e,f).
Albeit the difference in the signal amplitude, the overall shape of the color contrast is similar to that of Fig.~\ref{fig:color}(a-d), which suggests that the voltage reflects SAW induced spin waves at the Ni island.


\begin{figure}[bt]
	\includegraphics[scale=0.7]{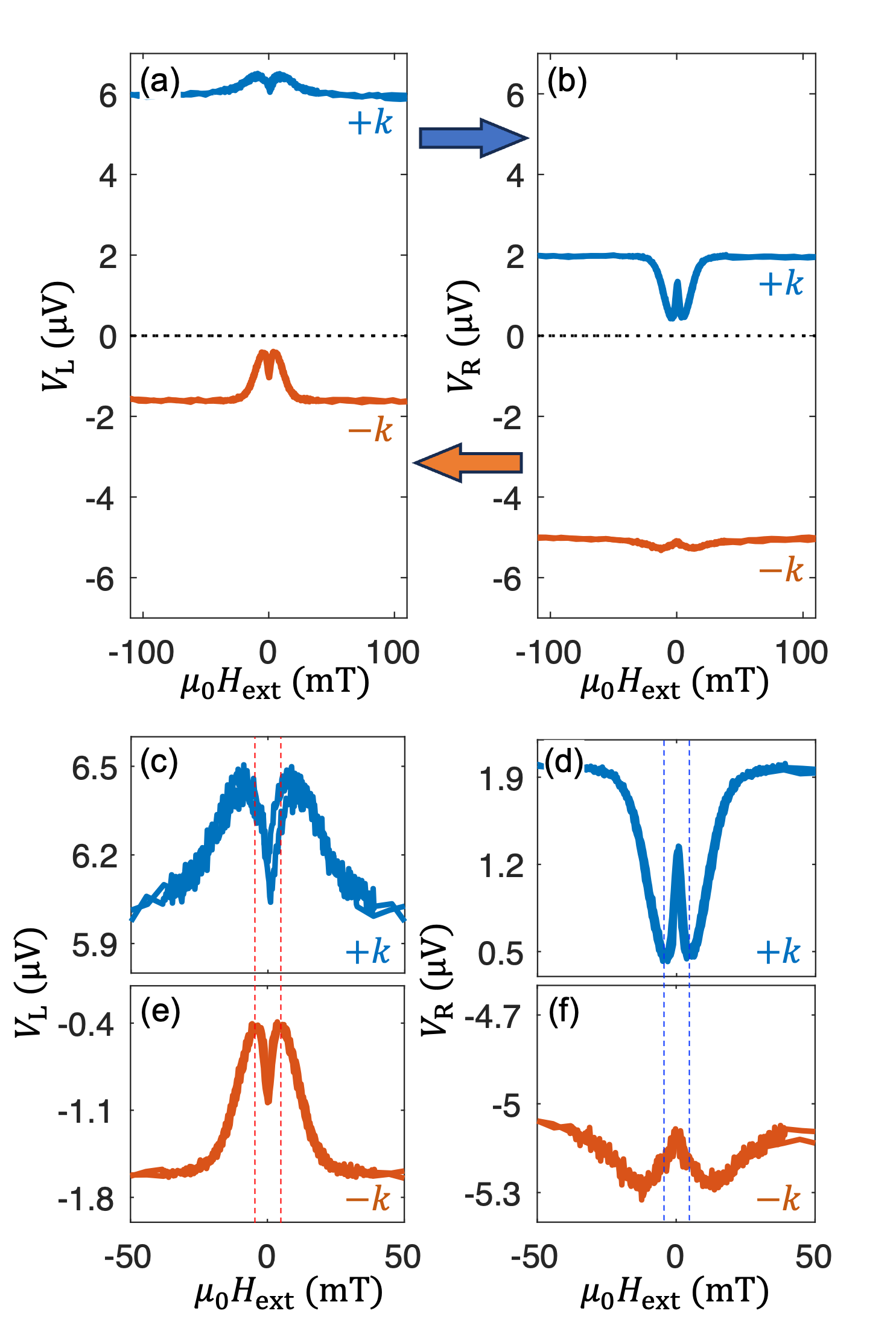}
	\caption{(a,b) $\mu_{0}H_\mathrm{ext}$ dependence of $V_\mathrm{L}$ (a) and $V_\mathrm{R}$ (b). Blue (red) lines: Measured voltage when an rf signal is applied to IDT1 (IDT2) and the SAW propagates along $+k$ ($-k$). (c-f) Expanded view of the data shown in (a,b). The vertical dashed lines are aligned with the spin wave resonance field obtained from the transmitted SAWs, i.e. the positions of the peaks in $V_\mathrm{L}$ for $-k$ and the dips in $V_\mathrm{R}$ for $+k$,.
	The external magnetic field is applied along $\varphi_\mathrm{H}$ = 60$^\mathrm{o}$.
	\label{fig:resonance}}
\end{figure}

To elaborate on this effect in more detail, we plot the $\mu_{0}H_\mathrm{ext}$ dependence of $V_\mathrm{L}$ and $V_\mathrm{R}$ with a fixed field angle ($\varphi_\mathrm{H} = 60^\mathrm{o}$) in Figs.~\ref{fig:resonance}(a) and \ref{fig:resonance}(b), respectively.
The blue (red) lines show data for $+k$ ($-k$).
Note that here we show voltages that are not normalized (the offset voltage is also shown).
Let us compare the voltages from the two antennas placed before ($V_\mathrm{L}$) and after ($V_\mathrm{R}$) the Ni island when the SAW propagates along $+k$ (i.e. the blue lines of Fig.~\ref{fig:resonance}(a,b)). 
First, the voltage at large $|\mu_{0}H_\mathrm{ext}|$ is larger for $V_\mathrm{L}$ that $V_\mathrm{R}$.
This is simply caused by the mechanical/electrical absorption of the SAW power at the Ni island; see the discussion pertaining to Fig.~\ref{fig:aec}(b).
Second, in contrast to the voltage dips found at $\mu_{0}H_\mathrm{res}$ for $V_\mathrm{R}$, we find a peak structure in $V_\mathrm{L}$.
These features can be understood if one considers the following scenario. 
The SAW launched at the IDT travels toward the Ni island. A fraction of the SAW that reach the island are reflected due to the difference in the mechanical/electrical properties of the substrate/air and substrate/Ni interfaces.
Since we apply a continuous wave rf signal to the IDT, incident and reflected SAWs coexist in the path along the IDT-Ni island and thus in the antenna.
The acoustoelectric current associated with the incident SAW (traveling toward the Ni island) in the antenna is the source of the large offset voltage found in Fig.~\ref{fig:aec}(a) and does not depend on the magnetic field since the incident SAW predominantly represents states before the SAW arrives at the Ni island.
In contrast, the reflected SAW carries information on SAW-induced spin wave excitation at the Ni island, similar to the transmitted SAW.
Note that the direction to which the acoustic current flows in the antenna is defined by the SAW propagation direction. 
Thus, the signal associated with SAW-induced spin wave excitation for the reflected SAW becomes opposite to that for the transmitted SAW (see e.g. Fig.~\ref{fig:aec}(b)).
The voltage measured at the antenna is the net sum of contributions from the incident and reflected SAWs, which can account for the signal observed experimentally.
The same argument applies to the dip structure found in $V_\mathrm{R}$ for $-k$ (Fig.~\ref{fig:resonance}(b), red line).

Interestingly, we find the position and width of the peak/dip in the measured voltages for the transmitted and reflected SAWs are different. 
To show this more clearly, an expanded view around the peak or dip is shown in Fig.~\ref{fig:resonance}(c-f).
The vertical dashed lines are aligned to the spin wave resonance field obtained from the transmitted SAWs: i.e. the positions of the peaks in $V_\mathrm{L}$ for $-k$ and those of the dips in $V_\mathrm{R}$ for $+k$.
The corresponding spin wave resonance field obtained from the reflected SAWs, i.e. the positions of the peaks in $V_\mathrm{R}$ for $-k$ and the dips in $V_\mathrm{L}$ for $+k$, are shifted to a larger $|\mu_{0}H_\mathrm{ext}|$, suggesting that the excited spin waves have a different resonance frequency. 
The width of the peak or dip is also larger for the spectra obtained with the reflected SAWs.

These results indicate that the reflected SAWs carry information of spin waves which is different from that of SAWs traversing the Ni island.
Given that the spin wave resonance frequency increases with increasing spin wave resonance field for the condition under study, the results in Fig.~\ref{fig:resonance}(c-f) show that spin waves with larger eigenfrequency are excited when the SAW is reflected at the Ni island.
The larger peak/dip width for the reflected SAWs suggests excitation of spin waves with larger damping and/or excitation of multiple modes with different frequencies.
At the moment, it is not clear why the peak/dip structure is different for the reflected and transmitted SAWs.
Note that the antenna used here allows one to probe SAW with frequency different from the SAW resonant (excitation) frequency. This is in stark contrast to the IDTs, which can only detect, in principle, integer multiples of the resonant frequency.
Thus the antenna serves as a broadband power detector of excited SAWs, similar to the power detectors in microwaves and optics.

In conclusion, we have demonstrated non-local detection of surface acoustic wave (SAW) driven spin waves using metallic antennas placed near a Ni island.
SAW induced voltage that develops at the antennas, which is proportional to the power of the traversing SAW, is measured. 
The voltage shows a magnetic field dependence that is almost identical to that of SAW transmission coefficient determined using a vector network analyzer.
The results suggest that the voltage measured at the antenna represents SAW power absorption due to SAW induced spin wave excitation.
The DC voltage detected at the antennas placed before and after the Ni island along the SAW propagation path suggests that the frequency and the linewidth of the spin waves excited by the reflected and transmitted SAWs at the island are different. 
Further study is required to identify the exact mechanism that causes such difference.
These results show the potential of metallic antennas as a simple power detector to study magnon-phonon coupling in magnetic nanostructures.

\section*{Acknowledgrements}
We are grateful to T. Funato, M. Matsuo, H. Komiyama, K. Taga, R. Hisatomi, Y. Shiota and S. Nakatsuji for fruitful discussion.
This work was supported by JSPS KAKENHI (Grant Number 20J20952, 20J21915, 23KJ1159, 23KJ1419, 23H05463) from JSPS, MEXT Initiative to Establish Next-generation Novel Integrated Circuits Centers (X-NICS) (Grant Number JPJ011438), and JSR Fellowship, the University of Tokyo.

\bibliography{ref_matsumoto}

\end{document}